\documentclass{pasj00}
\draft

\begin{document}

\SetRunningHead{Lei et al. }{Effect of binary fraction on HB morphology}


\title{Effect of Binary Fraction on Horizontal Branch Morphology
under Tidally Enhanced Stellar Wind}

\author{Zhenxin \textsc{Lei} }
\affil{Department of Science, Shaoyang University, Shaoyang 422000, China }
\affil{Key Laboratory for the Structure and Evolution of Celestial Objects,
Chinese Academy of Sciences,Kunming 650011,China}
\email{lzx2008@ynao.ac.cn}
\author{Xuemei \textsc{Chen} }
\affil{Department of Electrical Engeering, Shaoyang University, Shaoyang 422000, China }
\author{Xiaoyu {\sc Kang}, Fenghui {\sc Zhang} and Zhanwen {\sc Han}}
\affil{Yunnan Observatory,
  Chinese Academy of Sciences, Kunming 650011, China}
\affil{Key Laboratory for the Structure and Evolution of Celestial Objects,
Chinese Academy of Sciences,Kunming 650011,China}

%


\maketitle

\begin{abstract}
Tidally enhanced stellar wind may affect horizontal branch (HB)
morphology in globular clusters (GCs) by enhancing the mass loss of
primary star during  binary evolution. Lei et al. (2013a, 2013b) studied
the effect of this kind of wind on HB morphology in
details, and their results indicated that binary
is a possible second-parameter (2P) candidate in GCs.
Binary fraction is an very important fact in the tidally-enhanced-stellar-wind
model. In this paper, we studied the effect of binary fraction on HB morphology
by removing the effects of metallicity and age. Five different binary fractions
(i.e., 10\%, 15\%, 20\%, 30\% and 50\%)
are adopted in our model calculations. The
synthetic HB morphologies with different binary fractions are obtained at
different metallicities and ages.
We found that, due to the great influence of
metallicity and age, the effect of binary fraction
on HB morphology may be masked by these two parameters.
However, when the effects of metallicity and age
are removed, the tendency that HB morphologies become
bluer with increasing of binary fractions is clearly presented.
Furthermore, we compared our results
with the observation by Milone et al. (2012).  Our results are consistent
well with the observation at metal-rich and
metal-poor GCs. For the GCs with intermediate metallicity,
when the effect of age on HB morphology
is removed, a weak tendency that HB morphologies
become bluer with increasing of binary fractions is
presented in all regions of GCs, which is
consistent with our results obtained
in this metallicity range.
\end{abstract}

\section{Introduction}

The second parameter (2P) problem in globular clusters (GCs)
is a puzzling problem which challenges our understanding on
stellar evolution and formation history of GCs (see Catelan 2009
for a recent review). Most of the horizontal branch (HB) morphologies
in GCs can be described by their different metallicities
(Sandage \& Wallerstein, 1960). However, more and more evidence from
observations indicate that some other parameters, such as age, helium,
binary, planet system, cluster mass, etc are also need as the second or third parameter affecting
HB morphologies (Lee et al. 1994; Dotter et al. 2010;  Gratton et al. 2010;
D'Antona et al. 2002, 2005; D'Antona \& Caloi 2004, 2008;
Dalessandro et al. 2011, 2013; Valcarce, Catelan \& Sweigart 2012;
Lei et al. 2013a, 2013b; Soker 1998; Soker et al. 2000, 2001, 2007;
Recio-Blanco et al. 2006). However, among these parameters, none of
them can explain the whole HB morphologies in GCs successfully.
Recently, the 2P problem is considered
to be correlated with the multiple population phenomenon
(Piotto et al. 2007; Gratton, Carretta \& Bragaglia, 2012) and light elements anomalies
found in GCs (e.g., Na-O, Mg-Al anti-correlation, Marino et al. 2011, 2014;
also see Norris 1981; Norris et al. 1981; Kraft 1994),
which makes this puzzling problem become more complicated.

Lei et al. (2013a, 2013b, hereafter Paper I and Paper II respectively,
also see Han et al. 2012)
proposed that tidally enhanced stellar wind
during binary evolution may affect
HB morphology by enhancing the mass
loss of red giant primary star. In
this scenario, the different mass loss for the
progenitor of HB stars
is caused by different separations of binary systems,
and the number ratio of stars in different HB parts
is affected by binary fractions used in the models (see the
discussion in Paper II). These results
indicate that binary populations may
be the second or the third parameter candidate affecting
HB morphologies in GCs.
Sollima et al. (2007) and Milone et al. (2012)
estimated the binary fraction of many
Galactic GCs by analyzing the number
of stars located on the red side of
main-sequence (MS) fiducial line.
They found low binary fraction in
their GCs samples. Furthermore,
Milone et al. (2012) found small or
null relationship between binary fraction
and HB morphology in their observation samples. However,
they did not
remove the effects of other important parameters
when studying the effect of binary fraction on
HB morphology, such as metallicity and age.
Therefore, the motivation of this paper
is to study the effect of binary fraction
on HB morphology in details by considering the effects of
metallicity and age under the tidally enhanced
stellar wind.

The structure of this paper is as follows.
In Section~2, we introduce the tidally-
enhanced-stellar-wind model and code. The results
and comparison with observation are
given in Section~3. We discuss
our results in Section~4. Finally a conclusion
is given in Section~5.

\section{Models and Code}

The method used in this paper to obtain synthetic HB is
the same as in Paper I and Paper II.
Tidally enhanced stellar wind (Tout \& Eggleton 1988)
during binary evolution was incorporated into
Eggleton's stellar evolution code (Eggleton 1971, 1972, 1973;
see Paper I and Paper II for details) to
calculate the stellar mass and the helium core mass
of the primary star at the helium flash (hereafter
$M_{\rm HF}$ and $M_{\rm c,HF}$, respectively).
Then,  $M_{\rm HF}$, $M_{\rm c,HF}$, and a time spent on HB were used
to obtain the exact position of the primary star (e.g.,
effective temperature and luminosity)
on HB in the Hertzsprung-Russell
(H-R) diagram by interpolating  among constructed HB evolutionary tracks.
The time spent on HB in this paper means how long a HB star has been
evolved from zero-age HB (ZAHB). Since the
lifetime of HB stars is weakly dependent on
the total mass  when helium
core mass is fixed, the time spent
on HB is generated by a uniform random
number between 0 and $\tau_{\rm HB}$
(Rood 1973; Lee et al. 1990; Dalessandro et al. 2011).
Here, $\tau_{\rm HB}$ was set to be the lifetime of  HB star
with the lowest stellar mass
among the constructed HB evolutionary tracks,
which means
that this star  has the longest lifetime on the HB.
All the HB evolutionary tracks were constructed using
Modules for Experiments in Stellar Astrophysics (MESA;
Paxton et al. 2011; see Paper I for  details).
Finally, we transform the effective temperature and luminosity
of each HB star into $B-V$ colors and absolute magnitudes, $M_{\rm V}$,
using the Basel stellar spectra library (Lejeune et al. 1997, 1998)
to obtain the synthetic HB morphology in color-magnitude diagram (CMD).

The parameters adopted in Eggleton's  stellar evolution
code are the same as in Paper I and Paper II.
The tidally enhanced stellar wind during binary evolution
is described by the following equation,
\begin{equation}
\dot{M}=-\eta4\times10^{-13}(RL/M)\{1+B_{\rm w}\times \rm min [\it(R/R_{\rm L})^{6}, \rm 1/2^{6}]\},
\end{equation}
where $\eta$ is the Reimers mass-loss efficiency (Reimers 1975);
$R_{L}$ is the radius of the Roche lobe;
$B_{\rm w}$ is the tidal enhancement efficiency.
$R$, $L$, $M$ are the radius, luminosity and mass of the
primary star in solar units.

 \begin{table}
\renewcommand{\arraystretch}{0.8}

\centering

      \begin{minipage}[]{90mm}
   \caption{The input parameters for model calculations
   in this paper. The columns from left to right give
   metallicity, age of GCs and binary fraction respectively.}
   \end{minipage}\\

    \begin{tabular}{cc|c|cc}
    \hline
      &$Z$  & age (Gyr) &  $f_{\rm bin}$ (\%) \\
     \hline
      &     &10     &  10               \\
      &     &       &  15              \\
      &0.02 &12     &  20            \\
      &     &       &  30               \\
      &     &13     &  50              \\
      \hline\hline\noalign{\smallskip}
      &  &10       &  10             \\
      &  &         &  15              \\
      &0.001  &12  &  20              \\
      &  &         &  30               \\
      &  &13    &  50              \\
            \hline\hline\noalign{\smallskip}
      &  &10       &  10             \\
      &  &         &  15              \\
      &0.0001  &12 &  20              \\
      &  &         &  30               \\
      &  &13       &  50              \\

              \hline\noalign{\smallskip}

   \end{tabular}
      \end{table}

We generate several  groups of binary systems with different binary fractions.
The initial orbital periods of all
binary systems  are produced by
Monte Carlo simulations.
The distribution of
separation in binary is constant in $\log a$
($a$ is the separation) and falls off smoothly at small separations (Han et al. 2003),
\begin{equation}
a\cdot n(a)=\left\{
 \begin{array}{lc}
 \alpha_{\rm sep}(a/a_{\rm 0})^{\rm m}, & a\leq a_{\rm 0},\\
\alpha_{\rm sep}, & a_{\rm 0}<a<a_{\rm 1},\\
\end{array}\right.
\end{equation}
where $\alpha_{\rm sep}\approx0.07$, $a_{\rm 0}=10\,R_{\odot}$,
$a_{\rm 1}=5.75\times 10^{\rm 6}\,R_{\odot}=0.13\,{\rm pc}$ and
$m\approx1.2$.
To study the effect of binary fraction on HB morphology, we
adopt various value of binary fraction in model calculations
(i.e., 10\%, 15\%, 20\%, 30\% and 50\%)\footnote{Note that
these fractions are for the binary systems with
their orbital periods less than 100\,yr}.
The synthetic HB with different binary fractions
are obtained at different metallicities and ages.
The detailed input information
for our model calculations in this paper is given
in Table~1. The columns from left to right give
metallicity, age of GCs and binary fraction, respectively.
We use three different metallcities
(e.g., $Z$=0.02, 0.001, 0.0001) in our model
calculations, and for each metallicity,
we adopt three different ages (e.g., 10, 12 and 13 Gyr). For
each fixed metallicity and age,
we adopt five different binary fractions (e.g., 10\%,
15\%, 20\%, 30\% and 50\%).

\section{Results}

\begin{figure}
\centering
\includegraphics[width=90mm,angle=0]{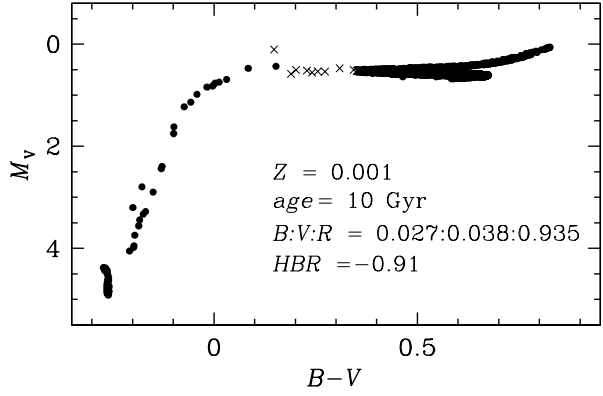}
\begin{minipage}[]{90mm}
\caption{An example of synthetic HB under tidally enhanced
stellar wind for a binary fraction of 30\%.
HB stars located in RR Lyrae instability strip
are denoted by crosses, and other HB stars are
denoted by solid dots. $B,V,R$ are the number of HB stars
bluer than (or to the left of), within and redder than
(or to the right of) the RR Lyrae instability strip. See the
text for details.} \end{minipage}
 \label{Fig1}
\end{figure}

\begin{figure}
\centering
\includegraphics[width=90mm,angle=0]{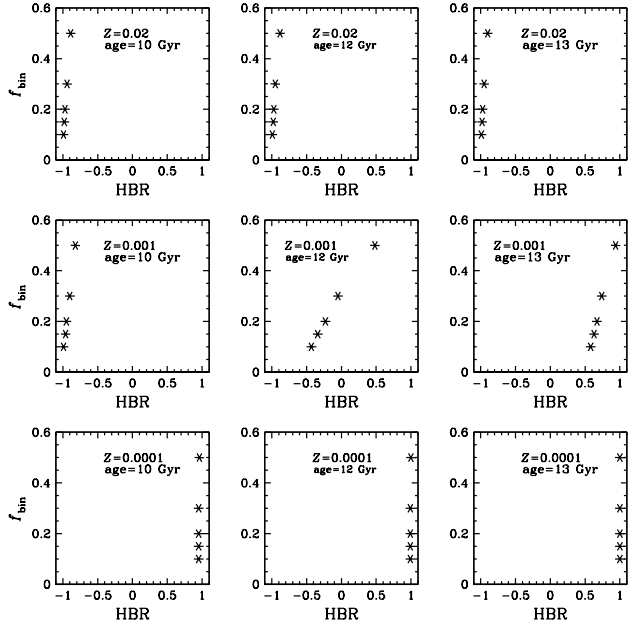}
\begin{minipage}[]{90mm}
\caption{Correlation between binary fraction and HB morphology
under tidally enhanced stellar wind. The metallicity and
age are fixed as labeled in each panel, but the binary fractions are different.} \end{minipage}
 \label{Fig2}
\end{figure}

Fig.1 shows an example of synthetic HB morphology under tidally enhanced stellar wind
for a binary fraction of 30\%.
In Fig.1, metallicity is $Z=0.001$; age is 10 Gyr.
HB stars located in RR Lyrae instability strip
are denoted by crosses, and other HB stars are
denoted by solid dots. The RR Lyrae instability strip is defined by
the vertical region of $3.80\leq$ log$T_{\rm eff}$ $\leq3.875$
in the H-R diagram (Koopmann et al. 1994; Lee et al. 1990,
see Fig.1 in Paper I).
The label, $B:V:R$, is the number ratio of stars
in different parts of HB, where $B,V,R$ are the number of HB stars
bluer than (or to the left of), within and redder than
(or to the right of) the RR Lyrae instability strip
(Lee et al. 1990). The label, $HBR$,
is a parameter to describe HB morphology of GCs (Lee et al. 1994),
and it is defined as follows,
\begin{equation}
HBR=(B-R)/(B+V+R),
\end{equation}
where $B, V, R$ have the same meaning as described above.
The value of $HBR$
is in the range of -1 to 1.
The value of -1 means that
all HB stars settle on red HB,
while the value of 1
means that GC presents a whole blue HB.
Therefore, the larger of the $HBR$ parameter,
the bluer of HB morphology in GCs.
For each synthetic HB morphology obtained in model
calculation listed in Table~1, we obtain
the value of $HBR$ to study the
correlation between binary fraction and
HB morphology in GCs. The results are
given in next section.

\subsection{Effect of Binary Fraction on HB Morphology}

Fig.2 shows the effect of binary fraction on HB morphology
under tidally enhanced stellar wind.
In each panel of Fig.2, the horizontal axis is
the parameter $HBR$ which is used to describe
HB morphology, and the vertical axis is
binary fraction.  The synthetic
GCs in each panel of Fig.2 have fixed metallcity and age but different
binary fractions.

One can see that for the three top panels of Fig.2, in which synthetic GCs have
a high metallicity of $Z$=0.02, though the binary fractions
are different (e.g., from 10\% to 50\% in each panel), all synthetic GCs
present a pure red HB morphology, and small or null effect of
binary fraction on HB morphology is revealed.
A similar result is also shown in the
three bottom panels of Fig.2, in which GCs have a very low
metallicity of $Z$=0.0001. All synthetic GCs in these three panels
show a pure blue HB morphology regardless of
the different binary fraction.
This is due to the fact that metallicity influence
HB morphology significantly at very high and very low
metallicity, and it may mask the effect of
binary fraction seriously.

\begin{figure}
\centering
\includegraphics[width=90mm,angle=0]{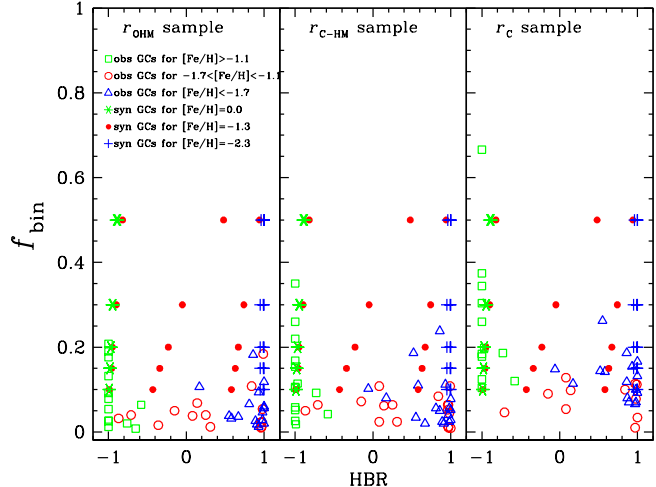}
\begin{minipage}[]{90mm}
\caption{Correlation between binary fraction and HB morphology
for different regions of GCs.
From left to right, these panels show the relationships
from outside region to core region of the GCs (see the
text for details). Green open squares, red open circles and
blue open triangles denote observed GCs in metallicity range of
[Fe/H]$\ >$-1.1, -1.7$\ <$ [Fe/H] $\ <$-1.1 and [Fe/H]$\ <$-1.7, respectively.
The green asterisks, red solid circles and
blue plus represent the synthetic GCs shown in Fig.2 at
metallicity of $Z$=0.02, 0.001, 0.0001
(or [Fe/H]=0.0, -1.3, -2.3) respectively.
 } \end{minipage}
 \label{Fig3}
\end{figure}

\begin{figure}
\centering
\includegraphics[width=80mm,angle=0]{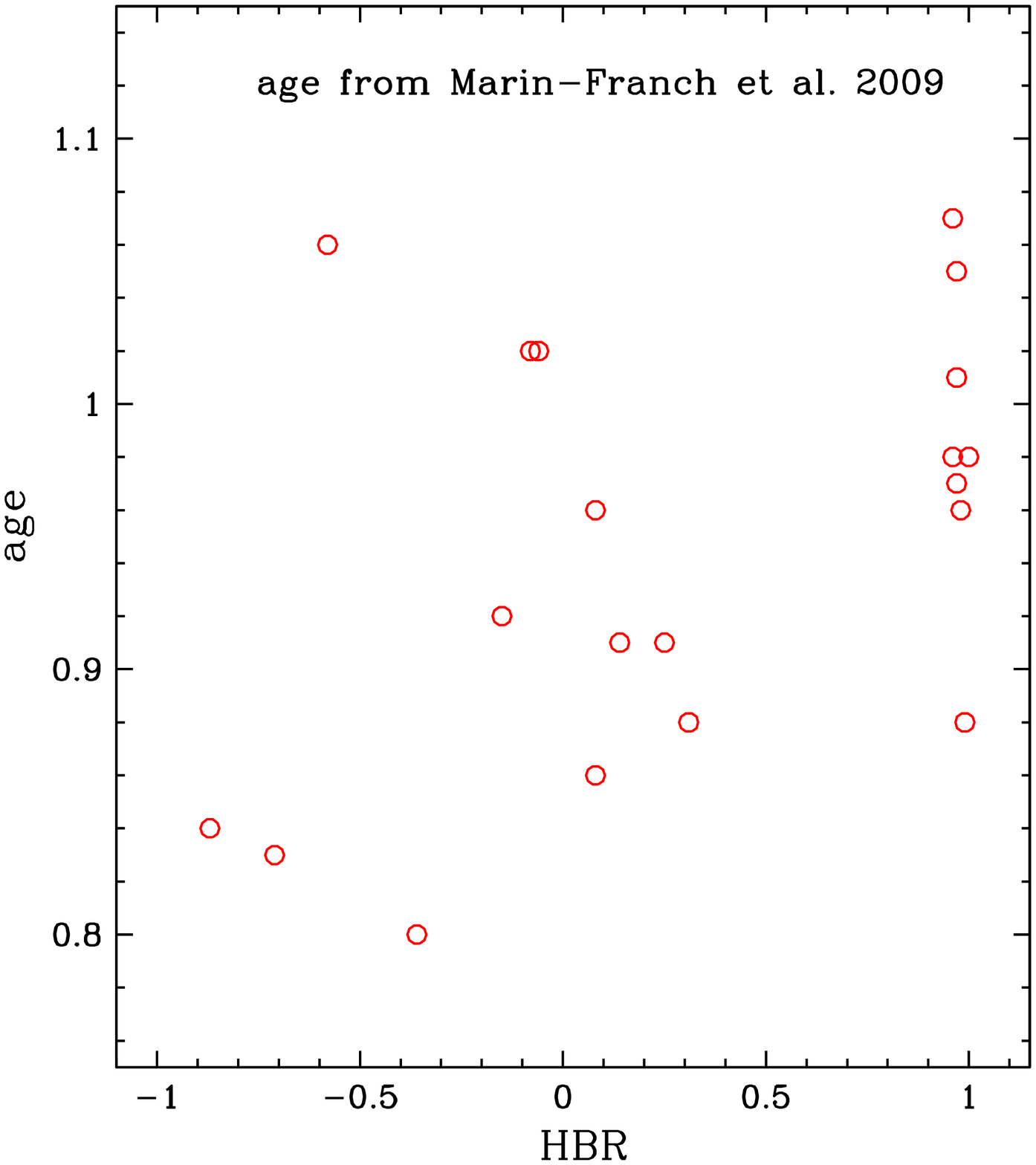}
\begin{minipage}[]{90mm}
\caption{Relationship between relative age and HB morphology for
GCs in the intermediate metallicity group (-1.7$\ <$ [Fe/H] $\ <$-1.1).  } \end{minipage}
 \label{Fig4}
\end{figure}

However, for the three middle panels of
Fig.2, in which synthetic GCs present an intermediate metallicity of $Z$=0.001,
one can see that with  the
binary fraction increasing, the synthetic
HB becomes bluer. This result is more clearly revealed
especially in the middle panel in which synthetic
GCs have an age of 12 Gyr (e.g., in this
panel, with the binary fraction increasing from 10\% to 50\%, the
value of $HBR$ increases from about -0.5 to 0.5, which means
the synthetic GCs transform from a dominant red HB to a blue HB).
Nevertheless, the correlation between binary fraction
and HB morphology is weakened in other two panels in which
GCs have ages of 10 and 13Gyr. This result indicates that,
age can also weaken the  effect of binary fraction
on HB morphology when GCs are very young
or very old.

The results obtained from Fig.2 indicate that
metallicity and age may mask the effect of binary
fraction on HB morphology. However,
for intermediate metallicity, when
the effect of age is considered, our results
reveal that higher binary fraction may make
HB morphology become bluer. This is due to the fact that, in our scenario, when
binary fraction increasing, more blue and extreme HB stars
will be produced under tidally enhanced stellar wind
(see the discussion in Paper II).  In the next section,
we will compare our results with recent observations.

\subsection{Comparison with Observations}

Milone et al. (2012) estimated the binary fraction
for 59 Galactic GCs by analyzing the
number of stars located on the red side of
main-sequence fiducial line. Furthermore, they studied the
relationship between binary fraction and HB morphology
of GCs and found that there is a small or null effect
of binary fraction on HB morphology. However, Milone
et al. (2012) did not consider the effects of metallicity
and age on HB morphology when studying the effect of
binary fraction. As we have seen in Fig.2, these two parameters
may mask the effect of binary fraction on HB morphology seriously.

In Fig.3, we compared our results obtained in Fig.2 with
the observation by Milone et al. (2012).
The binary fractions for observed GCs are from
Milone et al. (2012), while metallicity,
and HB morphology index, $HBR$, are from
Carretta et al. (2010).
The three panels
represent three different regions of GCs which
are defined by Milone et al. (2012).
$r_{\rm C}$ means the region within one core radius of GC, and
$r_{\rm C-HM}$ means the region between the core and
the half-mass radius in GCs, while $r_{\rm 0HM}$ means the
region outside the half-mass radius in GCs.
Therefore, the panels from left to right in Fig.3 show
the relationship between binary fraction and on HB morphology
from outside region to core region of GCs.

To weaken the effect of metallicity on
HB morphology, we divide the observed GCs into
three groups according to their metallicities.
They are metal-rich group ([Fe/H]$>$-1.1, denoted by
green open squares), intermediate metallicity group
(-1.7$\ <$[Fe/H]$\ <$-1.1, denoted by red open circles) and
metal-poor group ([Fe/H]$\ <$-1.7, denoted by blue open triangles)
respectively.
In each panel, green asterisks, red solid circles and
blue plus represent our synthetic GCs shown in Fig.2 at
metallicity of $Z$=0.02, 0.001, 0.0001 (or [Fe/H]=0.0, -1.3, -2.3),
which correspond to metal-rich, intermediate metallicity and
metal-poor group of observed GCs respectively
(note that, for each metallicity we adopt three different ages,
10, 12, 13 Gyr, see Fig.2).

One can see clearly that in each panel of Fig.3,
the observed GCs in metal-rich group show
red HB morphologies, while observed GCs in metal-poor
group present blue HB morphologies. We can not
find any correlations between binary fraction and
HB morphology for observed GCs in these two metallicity groups.
This is due to the fact that metallicity (the first parameter) can
significantly influence the HB morphology at very high and very low metallicity,
and it masks the effect of binary fraction
(even including the effect of age) on
HB morphology. In these two metallicity groups, our
synthetic GCs are consistent well with
the observational data (see green asterisks and blue plus in each panel
respectively). However, for the observed GCs in intermediate metallicity
group, at first glance, there is
still no evident correlation between
binary fraction and HB morphology
(see red open circles in each panel), which looks
like to contradict with our results for this intermediate metallicity group.
We will discuss this group of GCs in the next section in details.

\section{Discussion}

\begin{figure}
\centering
\includegraphics[width=90mm,angle=0]{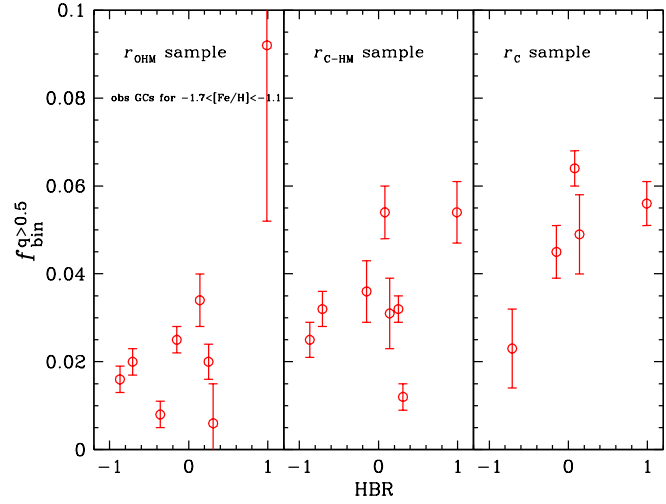}
\begin{minipage}[]{90mm}
\caption{Correlation between binary fraction
and HB morphology for GCs in the intermediate metallicity group
(-1.7$\ <$ [Fe/H] $\ <$-1.1), in which  the oldest GCs are removed. } \end{minipage}
 \label{Fig5}
\end{figure}

\begin{figure}
\centering
\includegraphics[width=90mm,angle=0]{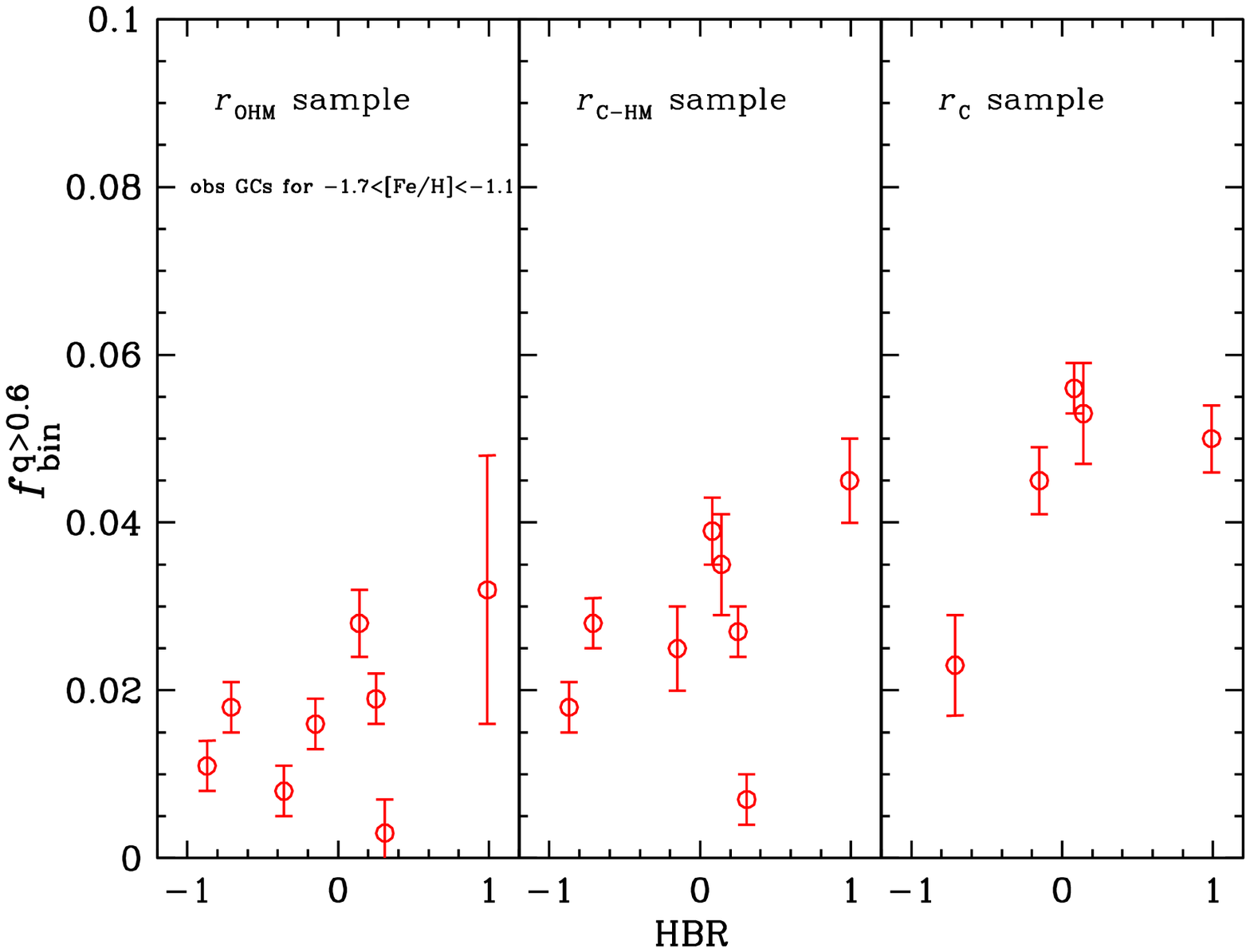}
\begin{minipage}[]{90mm}
\caption{Similar to Fig.5, but the binary fraction is for $q>$0.6. } \end{minipage}
 \label{Fig6}
\end{figure}

\begin{figure}
\centering
\includegraphics[width=90mm,angle=0]{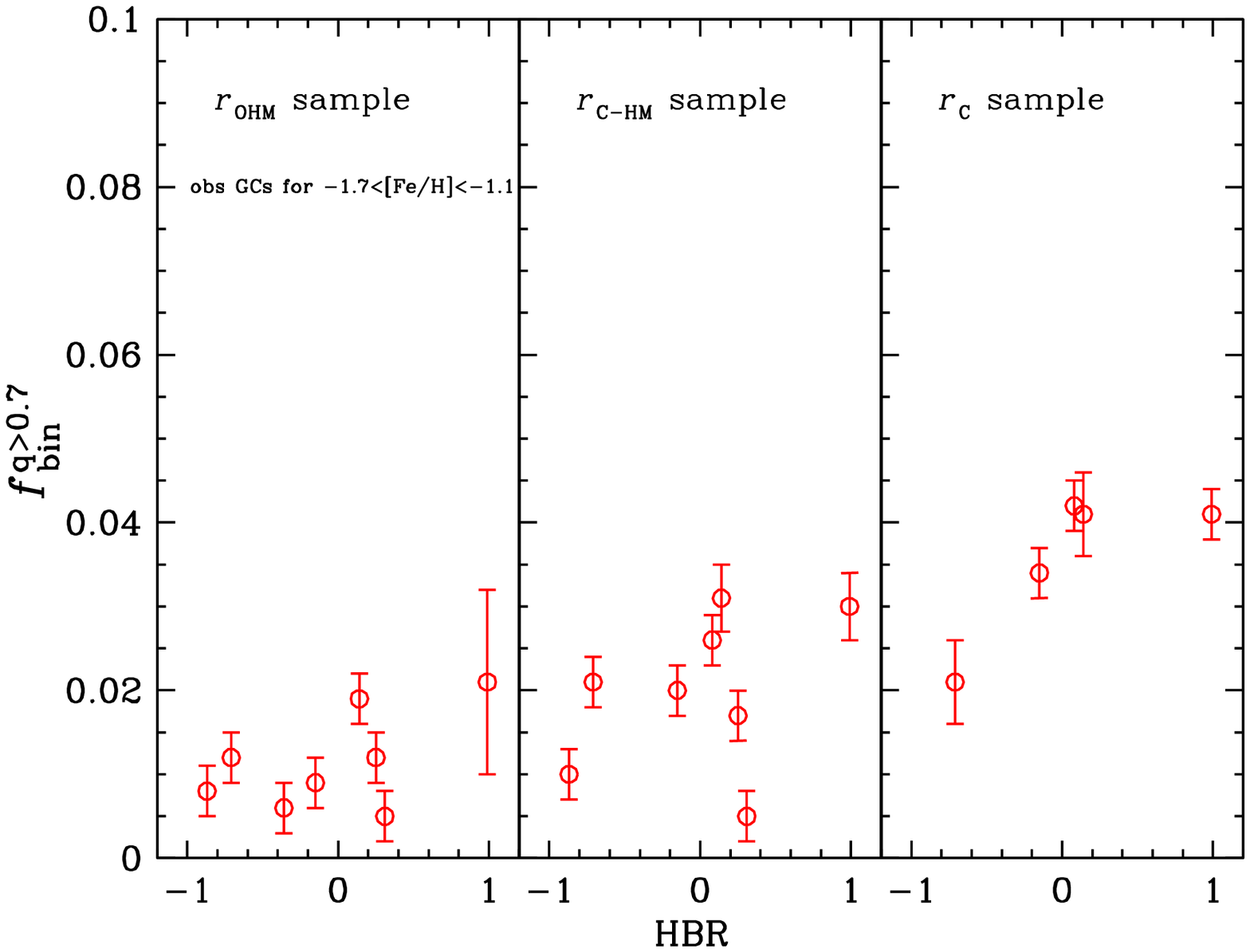}
\begin{minipage}[]{90mm}
\caption{Similar to Fig.5, but the binary fraction is for $q>$0.7. } \end{minipage}
 \label{Fig7}
\end{figure}

\begin{figure}
\centering
\includegraphics[width=90mm,angle=0]{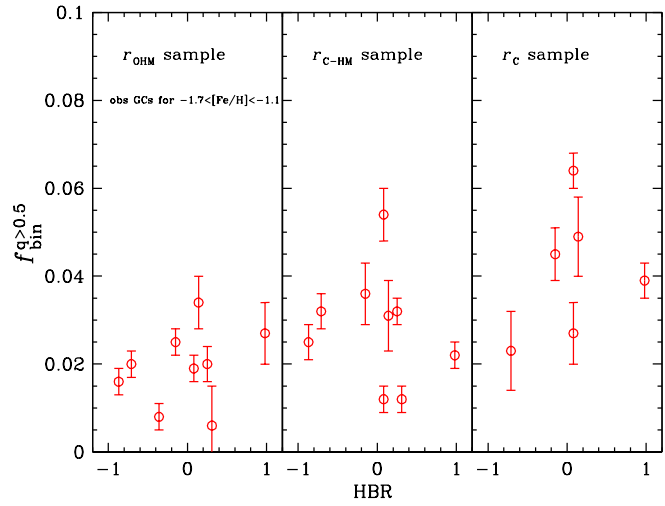}
\begin{minipage}[]{90mm}
\caption{Similar to Fig.5, but the age data of GCs is from
De Angeli et al. (2005)} \end{minipage}
 \label{Fig8}
\end{figure}

\begin{figure}
\centering
\includegraphics[width=90mm,angle=0]{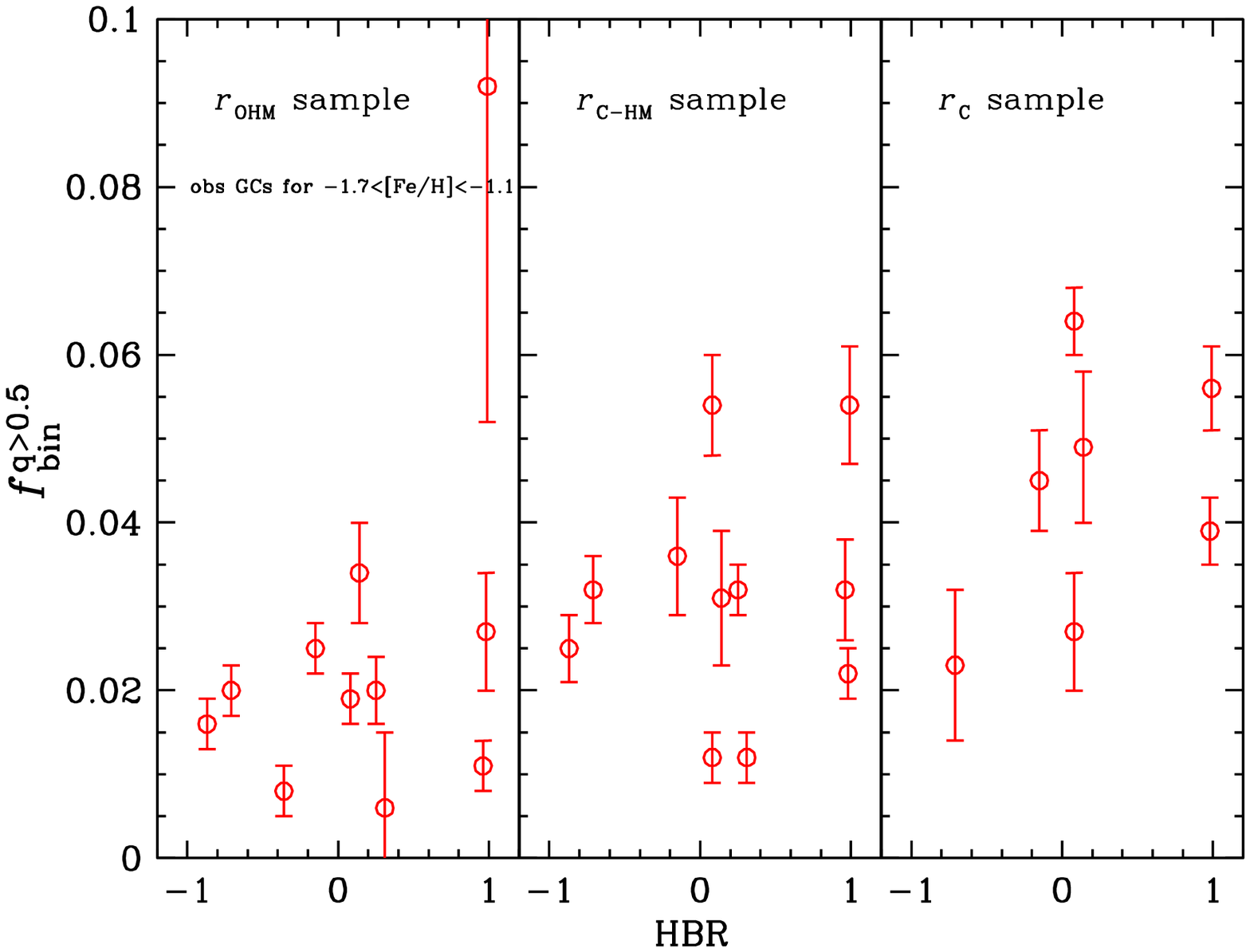}
\begin{minipage}[]{90mm}
\caption{Similar to Fig.5, but the age data of GCs is from
VandenBerg et al. (2013)} \end{minipage}
 \label{Fig8}
\end{figure}

Though the effect of metallicity on HB morphology is
weakened for observed GCs in the intermediate metallicity group,
the age of GCs still can influence the HB morphology
(Lee et al. 1994; Gratton et al. 2010; Dotter et al. 2010)
, and it may also mask the effect of binary fraction as we
discussed in Section~3.1.
For this reason, we studied the age of the observed GCs distributed in
the intermediate metallicity group.
Fig.4 shows the relationship between
relative age and HB morphology for observed GCs in
the intermediate metallicity group (-1.7$\ <$ [Fe/H] $\ <$-1.1).
The relative ages of the GCs are from
Marin-Franch et al. (2009) which are defined
as the ratio between cluster age and
the mean age of the clusters in low-metallicity group.
One can see clearly in Fig.4 that, the HB morphologies
become bluer with increasing age of GCs. Especially for
the oldest GCs (located in upper right of Fig.4),
their HB morphologies are significantly
affected by age, and
they present nearly a pure blue HB (e.g., $HBR\simeq$1).

To weaken the effect of
age on HB morphology,
We show in Fig.5 the relationship between binary fraction
and HB morphology for observed GCs in the intermediate metallicity group by
removing the oldest GCs (e.g., GCs with relative ages larger than about
0.96 in Fig.4 are removed).
Note that, the total binary fraction of GCs in Milone et al. (2012)
is obtained by assuming a constant mass-ratio
distribution between 0 and 1 (see Section 5.2 in Milone et al. 2012).
Under this assumption, it means that the binary fraction for $q>$0.5 is
equal to the binary fraction for $q<$0.5,
and the total binary fraction is simply 2
times of the binary fraction for  $q>$0.5.
Therefore, the total binary fractions
are dependent on the assumption of
mass-ratio distribution.
For this reason, the binary fraction used in
Fig.5 is for $q>$0.5, instead of the total binary fraction.
One can see in Fig.5 that, though not very obviously, all the
three panels present a weak correlation between binary
fraction and HB morphologies, especially in the core region.
With the binary fraction increasing, HB morphology becomes
bluer.

We also
study the effect of the
binary fractions for $q>$0.6 and 0.7 obtained in Milone et al. (2012)
on HB morphology in the intermediate
metallicity group of GCs (-1.7$<$ [Fe/H] $<$-1.1), and these binary fractions
are independent on the assumption of mass-ratio distribution.
The results are given in Fig.6 and Fig.7.
As in Fig.5,  we also weaken the
effect of age on HB morphology in these two figures by removing the oldest GCs
(e.g., GCs with their relative ages
larger than about 0.96 in Fig.4 are removed).
One can see that the relationship between binary fraction and HB
morphology present in Fig.6, 7 is very similar to the one in
Fig.5. The relative ages of GCs used in Fig.4 are from
Marin-Franch et al. (2009), and we also do the same analysis as in
Fig.5, 6, 7 but using the age data from De Angeli et al. (2005) and
VandenBerg et al. (2013) respectively. The results for $q>$0.5 are shown
in Fig.8 and 9, which are very similar to the one in Fig.5.

One may be aware of that the value of binary fractions adopted in our
model calculations (i.e., 10\%, 15\%, 20\%, 30\% and 50\%) are
different from the ones obtained by Milone et al. (2012, most
of them are less than 10\%). Note that, the binary fractions in
our model calculations are for the binary systems with
their  orbital periods less that 100 yr, but we do not
know any information of orbital periods for the binary fraction obtained in
Milone et al. (2012). Due to this reason, it is
unlikely to directly use the binary fraction of a certain
GC obtained by Milone et al. (2012)
as input in our model
calculations and then compare
the results with the observation.
However, the tendency that HB morphologies
of GCs become bluer with increasing of
binary fraction, though not very obviously in the observation\footnote{
This may be due to that binary is not the only second or third
parameter affecting HB morphology (e.g., helium, cluster mass, etc may also
at work in some GCs.
Freeman \& Norris 1981; Dotter et al. 2010;
Gratton et al. 2010). In different GCs, the dominate
second parameter may be different.},
is consistent with our model calculation when the
effects of metallicity and age are considered.

In a recent paper, Milone et al. (2014) defined
two new parameters to describe the HB morphology,
namely \emph{L}1 and \emph{L}2. They found that \emph{L}1,
which is the color difference between red giant branch (RGB) and HB in
the CMD of GCs, correlates with cluster age and metallicity, while
\emph{L}2, which is the color extension of HB, correlates with
the luminosity of GCs ($\it M_{\rm V}$) and helium abundance. 
Milone et al. (2014) also found an anti-correlation between
\emph{L}2 and binary fraction in group 2 and group 3 GCs (see Fig.10 in their paper),
which seems to demonstrate that a higher binary fraction corresponds to
a shorter HB extension. This result does not contradict with our results
obtained in this paper.
One can see from Fig.3 in Milone et al. (2014) that
more luminous (i.e. more massive) GCs present more extensional HB\footnote{A possible explanation is that more massive GCs can retain
the polluted material from the ejection of first-generation stars better than
the less massive GCs (Recio-Blanco et al. 2006). These polluted material
with higher helium abundance form the second-generation stars and extend the HB
into bluer region (D'Antona et al. 2002).}, hence a
longer \emph{L}2.
Meanwhile, more luminous (i.e. more massive) GCs also present
lower binary fractions (see Fig.40 in Milone et al. 2012)\footnote{ A possible
reason is that the mass of GCs and the binary destruction efficiency depend on
the cluster density and velocity dispersion
in a same way (Sollima 2008; Fregeau et al. 2009).}. 
These results reveal that the anti-correlation between \emph{L}2
and binary fraction could be a result of the correlation between \emph{L}2
and cluster mass and the anti-correlation between binary fraction and cluster mass.
Furthermore, the \emph{L}2 parameter describes the
extension of HB other than the specific number of HB stars in different HB regions.
That is why \emph{L}2 is sensitive to the $\emph{L}_{\rm t}$ parameter 
(Fusi-Pecci et al. 1993) and the maximum effective 
temperature of HB ($T_{\rm eff,Max}$; Recio-Blanco et al. 2006), while
insensitive to $HBR$ parameter (see Fig.12 in Milone et al. 2014) which depends on the
number ratio of HB stars located on different parts of HB. On the other hand,
in our binary scenario, different binary fractions would alter the
number of HB stars in different HB regions (see Paper I and Paper II for details),
hence it correlates with $HBR$ parameter.

\section{Conclusion}

In this paper, we studied the
effect of binary fraction on HB
morphology under tidally enhanced
stellar wind. We adopted five different
binary fractions, i.e., 10\%, 15\%, 20\%, 30\% and 50\%,  in our
model calculations.
Our results revealed that for the GCs with
intermediate metallicity and age, the correlation
between binary fraction and HB morphology is clear.
With binary fractions increasing, HB morphologies
become bluer. However,
metallcity and age may mask the effect of
binary fraction on HB morphology.
We also compared our results with the
observation of Milone et al. (2012).
We found that our results are consistent
well with the observed GCs in metal-rich
and metal-poor group. Moreover,
when the effect of
age on HB morphology for the intermediate
metallicity group of GCs is removed, a weak correlation
between binary fraction and HB morphology is
presented in all regions of GCs, and the
tendency that HB morphology becomes bluer with
increasing of binary fractions is consistent with our
results.

\bigskip

This work is supported by the Key Laboratory for
the Structure and Evolution of Celestial Objects, Chinese Academy of Science
(OP201302).



\begin{thebibliography}{}


\bibitem[{Carretta}{et~al.}(2010)]{Carretta10} Carretta, E., 
Bragaglia, A., Gratton, R. G.; et al. 2010, \aap, 516, A55

\bibitem[{Catelan}(2009)]{Catelan+09} Catelan, M. 2009, \apss, 320, 261


\bibitem[{Dalessandro}{et~al.}(2011)]{Dalessandro11} Dalessandro, E., Salaris, M.,
 Ferraro, F. R., et al. 2011, MNRAS, 410, 694

\bibitem[{Dalessandro}{et~al.}(2013)]{Dalessandro13} Dalessandro, E., Salaris, M.,
 Ferraro, F. R., Mucciarelli, A., Cassisi, S., 2013, MNRAS, 430, 459

\bibitem[{D'Antona}{et~al.}(2005)]{D'Antona05} D'Antona, F., Bellazzini, M., Caloi, V., Fusi Pecci, F., Galleti, S., \& Rood, R. T. 2005, \apj, 631, 868

\bibitem[{D'Antona \& Caloi}(2004)]{D'Antona04} D'Antona, F., \& Caloi, V. 2004, \apj, 611, 871

\bibitem[{D'Antona \& Caloi}(2008)]{D'Antona08} D'Antona, F., \& Caloi, V. 2008, MNRAS, 390, 693

\bibitem[{D'Antona}{et~al.}(2002)]{D'Antona02} D'Antona, F., Caloi, V., Montalb$\acute{a}$n, J., Ventura, P., \& Gratton, R. 2002, \aap, 395, 69

\bibitem[{De Angeli}{et~al.}(2005)]{De Angeli05} De Angeli, F.,  Piotto, G., Cassisi, S. et al.  2005, AJ, 130, 116

\bibitem[{Dotter}{et~al.}(2010)]{Dotter10} Dotter, A., Sarajedini, A., Anderson, J., et al. 2010, \apj, 708, 698

\bibitem[{Eggleton}(1971)]{Eggleton71} Eggleton, P. P. 1971, MNRAS, 151, 351

\bibitem[{Eggleton}(1972)]{Eggleton72} Eggleton, P. P. 1972, MNRAS, 156, 361

\bibitem[{Eggleton}(1973)]{Eggleton73} Eggleton, P. P. 1973, MNRAS, 163, 279

\bibitem[{Freeman \& Norris}(1981)]{Freeman81} Freeman, K. C. \& Norris, J. 1981, ARA\&A, 19, 319

\bibitem[{Fregeau}(2009)]{Fregeau09} Fregeau, J. M., Ivanova, N., \& Rasio, F. A. 2009, \apj, 707, 1533

\bibitem[{Fusi Pecci}{et~al.}(1993)]{Fusi Pecci93} Fusi Pecci, F., Ferraro, F. R., Bellazzini, M.,
Djorgovski, S., Piotto, G., \& Buonanno, R. 1993, AJ, 105, 1145

\bibitem[{Gratton}{et~al.}{2012}]{Gratton12} Gratton, R. G., Carretta, E., \& Bragaglia, A., 2012, \aapr, 20, 50

\bibitem[{Gratton}{et~al.}{2010}]{Gratton10} Gratton, R. G., Carretta, E., Bragaglia, A., et al. 2010, \aap, 517, A81

\bibitem[{Han}{et~al.}{2012}]{Han12} Han, Z., Chen, X., Lei, Z.,
\& Podsiadlowski, P. 2012, ASPC, 452, 3

\bibitem[{Han}{et~al.}{2003}]{Han03} Han, Z., Podsiadlowski, Ph., Maxted, P. F. L., \& Marsh, T. R. 2003, MNRAS, 341, 669

\bibitem[{Koopmann}{et~al.}(1994)]{Koopmann94} Koopmann, R. A., Lee, Y. W., Demarque, P., et al.   1994, \apj, 423, 380

\bibitem[{Kraft}(1994)]{Kraft94} Kraft, R. P., 1994, PASP, 106, 553

\bibitem[{Lee}{et~al.}(1990)]{Lee90} Lee, Y. W., Demarque, P., \& Zinn, R. 1990, \apj, 350, 155

\bibitem[{Lee}{et~al.}(1994)]{Lee94} Lee, Y. W., Demarque, P., \& Zinn, R. 1994, \apj, 423, 248

\bibitem[{Lei}{et~al.}(2013)]{Lei13a} Lei, Z., Chen, X.,
Zhang, F., \& Han, Z. 2013a, \aap, 549, A145, Paper I

\bibitem[{Lei}{et~al.}(2013)]{Lei13b} Lei, Z.,
Zhang, F., Ge, H., \& Han, Z. 2013b, \aap, 554, A130, Paper II

\bibitem[{Lejeune}{et~al.}(1997)]{Lejeune97} Lejeune, T., Cuisinier, F.,  \&  Buser R. 1997, \aap S, 125, 229

\bibitem[{Lejeune}{et~al.}(1998)]{Lejeune98} Lejeune, T., Cuisinier, F.,  \&  Buser R. 1998, \aap S, 130, 65

\bibitem[{Marin-Franch}{et~al.}(2009)]{Marin09} Marin-Franch, A.,
Aparicio, A., \&  Piotto, G., et al. 2009, \apj, 694, 1498

\bibitem[{Marino}{et~al.}(2014)]{Marino14} Marino, A. F.,
Milone, A. P., Przybilla, N., et al. 2014, MNRAS, 437, 1609

\bibitem[{Marino}{et~al.}(2011)]{Marino11} Marino, A. F., Villanova, S.,
Milone, A. P., et al. 2011, \apj, 730, L16

\bibitem[{Milone}{et~al.}(2014)]{Milone14} Milone, A. P.,
Marino, A. F., Dotter, A., et al. 2014, \apj, 785, 21

\bibitem[{Milone}{et~al.}(2012)]{Milone12} Milone, A. P.,
Piotto, G., Bedin, L. R., et al. 2012, \aap, 540, A16

\bibitem[{Norris}(2012)]{Norris81a} Norris, J. 1981, \apj, 248, 177

\bibitem[{Norris}{et~al.}(1981)]{Norris81b} Norris, J., Cottrell, P. L.; Freeman, K. C.; Da Costa, G. S. \apj, 1981, 244, 205

\bibitem[{Paxton}{et~al.}(2011)]{Paxton11} Paxton, B., Bildsten, L., Dotter, A., et al. 2011, ApJS, 192, 3

\bibitem[{Piotto}{et~al.}(2007)]{Piotto07} Piotto, G., Bedin, L. R., Anderson, J., et al.  2007, \apj, 661, L53


\bibitem[{Recio-Blanco}{et~al.}(2006)]{Recio+06} Recio-Blanco, A., Aparicio, A., Piotto, G., De Angeli, F., \& Djorgovski, s. G. 2006, \aap, 452, 875

\bibitem[{Reimers}(1975)]{Reimers75} Reimers, D. 1975, MSRSL, 8, 369

\bibitem[{Rood}(1973)]{Rood73} Rood, R. T. 1973, \apj, 184, 815

\bibitem[{Sandage \& Wallerstein}(1960)]{Sandage1960} Sandage, A., \& Wallerstein, G. 1960, \apj, 131, 598

\bibitem[{Soker}(1998)]{Soker98} Soker, N. 1998, AJ, 116, 1308

\bibitem[{Soker \& Hadar}(2001)]{Soker+01} Soker, N., \& Hadar, R. 2001, MNRAS, 324, 213

\bibitem[{Soker \& Harpaz}(2000)]{Soker+00} Soker, N., \& Harpaz, A. 2000, MNRAS, 317, 861

\bibitem[{Soker \& Harpaz}(2007)]{Soker07} Soker, N., \& Harpaz, A. 2007, \apj, 660, 699

\bibitem[{Sollima}(2008)]{Sollima08} Sollima, A. 2008, MNRAS, 388, 307

\bibitem[{Sollima}{et~al.}(2007)]{Sollima07} Sollima, A., Beccari, G., Ferraro, F. R., et al. 2007, MNRAS, 380, 781

\bibitem[{Tout \& Eggleton}(1988)]{Tout88} Tout, C. A., \& Eggleton, P. P. 1988, MNRAS, 231, 823

\bibitem[{Valcarce, Catelan \& Sweigart}(2012)]{Valcarce12} Valcarce, A. A. R.,
Catelan, M., \& Sweigart, A. V., 2012, \aap, 547, A5

\bibitem[{VandenBerg}{et~al.}(2013)]{VandenBerg13} VandenBerg, Don A.; 
Brogaard, K.; Leaman, R.; Casagrande, L. 2013, ApJ, 775, 134

\end{thebibliography}
\end{document}